\documentclass[10pt, conference]{IEEEtran}
\IEEEoverridecommandlockouts
\usepackage{cite}
\usepackage{amsmath,amssymb,amsfonts}
\usepackage{mathtools}
\usepackage{algorithm,enumerate,multirow,tikz}
\usepackage{algpseudocode}
\usepackage{graphicx}
\usepackage{textcomp}
\usepackage{xcolor}
\usepackage{bm}
\usepackage{pgf}
\usepackage{layouts}
\usepackage{array}
\usepackage{multirow}
\usepackage{float}
\usepackage{booktabs}
\def\BibTeX{{\rm B\kern-.05em{\sc i\kern-.025em b}\kern-.08em
    T\kern-.1667em\lower.7ex\hbox{E}\kern-.125emX}}

\makeatletter
\def\endthebibliography{%
  \def\@noitemerr{\@latex@warning{Empty `thebibliography' environment}}%
  \endlist
}
\makeatother

\newcommand{\bB}{\ifmmode{\mathbb{B}}%
	\else{\hbox{$\mathbb{B}$}}\fi}
\newcommand{\bE}{\ifmmode{\mathbb{E}}%
	\else{\hbox{$\mathbb{E}$}}\fi}
\newcommand{\bF}{\ifmmode{\mathbb{F}}%
	\else{\hbox{$\mathbb{F}$}}\fi}
\newcommand{\bM}{\ifmmode{\mathbb{M}}%
	\else{\hbox{$\mathbb{M}$}}\fi}
\newcommand{\bN}{\ifmmode{\mathbb{N}}%
	\else{\hbox{$\mathbb{N}$}}\fi}
\newcommand{\bR}{\ifmmode{\mathbb{R}}%
	\else{\hbox{$\mathbb{R}$}}\fi}
\newcommand{\bS}{\ifmmode{\mathbb{S}}%
	\else{\hbox{$\mathbb{S}$}}\fi}

\newcommand{\Ftwo}{\ifmmode{\mathbb{F}_2}%
	\else{\hbox{$\mathbb{F}_2$}}\fi}
\newcommand{\SNR}{\ifmmode{E_\text{b}/N_0}%
	\else{\hbox{$E_\text{b}/N_0$}}\fi}

\newcommand\Tstrut{\rule{0pt}{2.4ex}}
\newcommand\Bstrut{\rule[-0.9ex]{0pt}{0pt}}
 
\newcommand{\proj}[3][]{%
    \ifthenelse{\equal{#1}{}}%
    {\ifmmode{#2_{/\mathbb{#3}}}\else{\hbox{$#2_{/\mathbb{#3}}$}}\fi}%
    {\ifmmode{#2_{/\mathbb{#3}_#1}}\else{\hbox{$#2_{/\mathbb{#3}_#1}$}}\fi}%
    }
\def\RM{\ifmmode{\mathcal{RM}(m,\,r)}%
	\else{\hbox{$\mathcal{RM}(m,\,r)$}}\fi}
\def\rm#1#2{\ifmmode{\mathcal{RM}(#1,\,#2)}%
	\else{\hbox{$\mathcal{RM}(#1,\,#2)$}}\fi}

\newtheorem{definition}{Definition}

\begin{document}

\title{Semi-Deterministic Subspace Selection for Sparse Recursive Projection-Aggregation Decoding of Reed-Muller Codes
\thanks{This work has received funding from the European Research Council (ERC) under the European Union’s Horizon 2020 research and innovation programme (grant agreement No. 101001899).}
}

\author{\IEEEauthorblockN{Johannes Voigt, Holger Jäkel and Laurent Schmalen}
\IEEEauthorblockA{Communications Engineering Lab (CEL), Karlsruhe Institute of Technology (KIT)}
Hertzstr.~16, 76187 Karlsruhe, Germany, Email: \texttt{\{johannes.voigt, holger.jaekel\}@kit.edu}
}

\maketitle
\begin{abstract}\label{sec:abstract}
Recursive projection aggregation (RPA) decoding as introduced in \cite{9023389} is a novel decoding algorithm which performs close to the maximum likelihood decoder for short-length Reed-Muller codes.
Recently, an extension to RPA decoding, called sparse multi-decoder RPA (SRPA), has been proposed \cite{9517887}.
The SRPA approach makes use of multiple pruned RPA decoders to lower the amount of computations while keeping the performance loss small compared to RPA decoding.
However, the use of multiple sparse decoders again increases the computational burden.
Therefore, the focus is on the optimization of sparse single-decoder RPA decoding to keep the complexity small.

In this paper, a novel method is proposed, to select subsets of subspaces used in the projection and aggregation step of SRPA decoding in order to decrease the decoding error probability on AWGN channels.
The proposed method replaces the random selection of subspace subsets with a semi-deterministic selection method based on a figure of merit that evaluates the performance of each subspace.
Our simulation results show that the semi-deterministic subspace selection improves the decoding performance up to $0.2\,\text{dB}$ compared to SRPA.
At the same time, the complexity of SRPA decoding for RM codes of order $r\geq 3$ is reduced by up to 81\% compared to SRPA.
\end{abstract}


\section{Introduction}\label{sec:introduction}
Reed-Muller (RM) codes are linear block codes developed almost 70 years ago by Muller \cite{6499441}. A first efficient decoding algorithm (Reed's algorithm) was later published by Reed \cite{1057465}.
Although RM codes are one of the oldest codes, they are perhaps more relevant than ever \cite{9123985}, since they are closely related to the widely used polar codes.
In 2017, it has been shown that RM codes achieve capacity on the binary erasure channel \cite{7862912}.
Furthermore, recently they have been shown to achieve capacity on general binary memoryless symmetric (BMS) channels \cite{2110-14631}.
Thus, new decoding algorithms for RM codes have been developed in recent years.

The recursive projection-aggregation (RPA) decoding algorithm, proposed by Ye and Abbe \cite{9023389} in 2020, received attention due to its decoding performance close to maximum likelihood (ML) decoding for short-length ($\leq 1024$ bits) RM codes.
Unlike ML decoding, RPA decoding is feasible for second- and higher-order RM codes.
Nevertheless, the computational complexity of RPA decoding is its main bottleneck and must be reduced for RPA to be used in practical applications.
The idea behind RPA is to recursively project RM codes onto its cosets to get first-order and shorter-length RM codes.
These are decoded efficiently using fast Hadamard transform (FHT) ML decoding.
Finally, the decoded projections are aggregated to reconstruct the originally RM code.
RPA decoding can outperform successive cancellation list (SCL) decoding of polar codes with cyclic redundancy check (CRC) \cite{9023389}.

To reduce the complexity of RPA, several extensions were proposed recently.
A reduced complexity RPA algorithm ($\text{RPA}_{\text{SYN}_\delta+\text{SCH}}$) with a syndrome-based early stopping technique and a scheduling scheme to reduce the number of iterations and projections has been proposed in \cite{9594060}.
In \cite{9174087}, a collapsed projection-aggregation (CPA) decoder has been published, which avoids the recursions of RPA by directly projecting to the first-order RM codes.
The pruned CPA (PCPA) decoder \cite{2105.11878} decreases the number of projections by pruning.
An optimized PCPA with simplifications of the projection and aggregation function has been proposed recently in \cite{9745165}.
In addition to these approaches, sparse multi-decoder RPA (SRPA) decoding \cite{9517887} reduces the computational complexity of RPA by up to 80\% by using only a random subset of projections, e.\,g., using 20\% of all subspaces, for decoding.
The number of projections used can be set by a parameter.
Unfortunately, a decreasing number of projections leads to a degradation of the word error rate (WER).
This can be compensated by using $k\in\bN$ SRPA decoders ($k$-SRPA).
Consequently, this results in a $k$-fold higher complexity, which reduces the benefit gained from pruning the projections.
The subsequent choice of the final codeword estimation can either be done by choosing the most likely reconstruction based on the received word or by using a CRC verification, reducing the effective code rate.
Based on SRPA, a multi-factor pruning approach further reduces the number of projections \cite{2208.13659}.

To fulfill practical energy consumption constraints, the decoder should be of low complexity while having a sufficiently good performance to achieve a required target WER.
All the mentioned approaches reduce the computational complexity of RPA decoding.
In particular, pruning the projections reduces the complexity effectively but comes with a compromise between performance and complexity.
For the pruning in SRPA, one-dimensional subspaces are chosen at random from a uniform distribution to generate the cosets used in the projection step.
In \cite{9517887}, it remains open if subspaces could be selected deterministically to lower the WER of SRPA decoding without increasing the complexity.

In this paper, we focus on the enhancement of low-complexity single-decoder SRPA for general BMS channels to improve the performance of SRPA decoding and reducing the complexity simultaneously.
We propose a novel method to select the subspaces used for the projections of SRPA instead of choosing them randomly.
The method is based on a figure of merit which evaluates the potential performance of the individual subspaces, depending on the received codeword.
To guarantee diversity in the subsets of subspaces in consecutive iterations of SRPA, we choose a random subset out of a deterministically selected subset of all subspaces.
We call this method semi-deterministic subspace selection (SDSS).

The simulation results presented below in Sec.\,\ref{sec:results} show that the SDSS method reduces the WER of SRPA decoding up to 20\%, corresponding to a coding gain by up to  $0.2\,\text{dB}$, depending on the RM code and the pruning factor used.
In addition, the computational burden for decoding RM codes of order $r\geq 3$ is effectively reduced by the proposed optimization techniques.
\section{Preliminaries}\label{sec:preliminaries}

\subsection{Reed-Muller Codes}\label{subsec:rm-code}
Binary RM codes can be defined using Boolean functions. 
Let $f(v_1,\ldots,v_m): \bF_2^m \rightarrow \Ftwo$ be a Boolean function in $m$ variables.
The binary variables $v_1,\ldots,v_m$ are represented through a vector $\bm{z}\in \bE \coloneqq \bF^m_2$, where $v_1$ represents the least-significant bit of $\bm{z}$ and $v_m$ represents the most-significant bit accordingly.
The evaluation of $f$ on each $\bm{z}\in \bE$ yield a unique vector $\bm{f}\in\bF^{2^m}$ representing $f$.
One way to index the coordinates of $\bm{f}$ is via the vectors $\bm{z}\in\bE$ introduced previously, following:
\begin{equation}\label{eq:indexing-c}
\bm{f} = \big( f(\bm{z}_i), \bm{z}\in\bE \big).
\end{equation}
Since $\bm{f}$ is of length $n=2^m$, there exist $2^n$ unique Boolean functions in $m$ variables.
The set of all Boolean functions in $m$ variables forms a vector space $V$ with the set of monomials
\begin{equation*}
    M = \{1, v_1, v_2, \ldots ,v_m, v_1v_2,\ldots, v_{m-1}v_m,\ldots, v_1v_2\cdots v_m\}
\end{equation*}
as a basis.
Each function in $V$ can be represented as a linear combination of the monomials in $M$.
With this, RM codes can be defined as follows:
\begin{definition}[Binary Reed-Muller Code] \cite[p.\,151]{wicker1995error}
    The $r$-th order binary Reed-Muller code, denoted as  \rm{m}{r}, consists of all linear combinations of vectors $\bm{f}$ associated with Boolean functions $f$ that are monomials of degree $0,1,\ldots,r$ in $m$ variables. The dimension of this code is $k=\sum_{i=0}^r \binom{m}{i}$, the code length is $n=2^m$ and $0\leq r \leq m$ must hold.
\end{definition}

\subsection{Recursive Projection-Aggregation Decoding}\label{subsec:rpa}
Since we focus on general BMS channels, defined as ${\text{W}:\Ftwo \rightarrow \bR}$, the log-likelihood ratio (LLR) of a channel output symbol $y\in\bR$:
\begin{equation}\label{eq:llr}
    \text{LLR}(y) \coloneqq \ln \left( \frac{\text{W}(y|c=0)}{\text{W}(y|c=1)}\right),
\end{equation}
where $c\in\Ftwo$ is the channel input symbol.
Following the notation in \cite{9023389}, we use the vectors $\bm{z}\in\bE$ to index the coordinates of $\bm{y}\in\bR^n$, such that $\bm{y}(\bm{z})\in\bR$ is the $\bm{z}$-th coordinate of $\bm{y}$.
Further, $\bm{L}\in\bR^n$ denotes the LLR vector of $\bm{y}$ and we write \eqref{eq:llr} as $\bm{L}(\bm{z})\coloneqq \text{LLR}(\bm{y}(\bm{z}))$ for short. %

The RPA algorithm \cite{9023389} consists of three main steps: the recursive projection of the received codeword, the decoding of the projections, and the aggregation of the reconstructions.
Let $\bm{c}\in\rm{m}{r}$ denote a codeword of an $r$-th order RM code of length $n=2^m$.
The coordinates of $\bm{c}$ can be indexed by the vectors $\bm{z}\in \bE$ following \eqref{eq:indexing-c}.
Moreover, let $\bB\subseteq\bE$ be an $s$-dimensional subspace and $s\leq r$.
Then the quotient space $\bE/\bB$ consists of all cosets of the form $T = \bm{z} + \bB$ for $\bm{z}\in\bE$.
In the following, we only use one-dimensional subspaces $\bB$, i.\,e., $s=1$.
Thus, there exist exactly $n-1$ non-zero one-dimensional subspaces $\bB$ with $n/2$ cosets $T \in \bE/\bB$ per subspace and $2$ unique elements $\bm{z}\in\bE$ each.

\subsubsection{Projection}\label{subsec:rpa-projection}
A received vector $\bm{y}\in\bR^n$, or its LLR vector $\bm{L}\in\bR^n$ respectively, can now be projected using the $n-1$ subspaces. The BMS projection function is defined as \cite{9023389}:
\begin{equation}\label{eq:rpa-awgn-prj}
    \proj{\bm{L}}{B}(T) = \ln \Big( \exp \big(\sum  _{\bm{z}\in T} \bm{L}(\bm{z}) \big) + 1 \Big) - \ln \Big(\sum_{\bm{z}\in T}\exp\big(\bm{L}(\bm{z})\big) \Big).
\end{equation}
In \eqref{eq:rpa-awgn-prj}, two coordinates of $\bm{L}$ indexed by the vectors $\bm{z}\in T \in \bE/\bB$ are combined.
Doing this for all $T \in \bE/\bB$ results in a new vector \proj{\bm{L}}{B} of length $n/2$, called projection.
Since there are $n-1$ one-dim subspaces $\bB$, this results in $n-1$ projections of length $n/2$.
These projections are RM codes of order $r-1$ and length $2^{m-1}=n/2$ \cite{9023389}.
The projection step is recursively repeated $r-1$ times until first-order RM code \rm{m-r+1}{1} projections have been reached.

\subsubsection{Decoding the Projections}\label{par:fht}
Decoding the first-order RM codes at the lowest-level recursion can efficiently be done using fast Hadamard transform (FHT) decoding \cite{1057189}.
The FHT is a soft-decision maximum likelihood (ML) decoder with complexity $\mathcal{O}(n \log n)$, compared to $\mathcal{O}(n^2)$ of a naive ML decoder \cite{9123985}.
The algorithm is given in \cite[p.\,381]{10.5555/1088886}.
The FHT decoder takes the projections $\proj{\bm{L}}{B}\in\bR^{2^{m-r+1}}$ and returns the reconstructions $\proj{\hat{\bm{y}}}{B} \in \bF_2^{2^{m-r+1}}$.

\subsubsection{Aggregation}\label{par:aggregation}
To obtain the estimate $\hat{\bm{L}}$ of $\bm{L}$, the reconstructions are aggregated recursively using the function \cite{9023389}:
\begin{equation}\label{eq:rpa-awgn-agg}
    \hat{\bm{L}}(\bm{z}) = \frac{1}{n-1} \sum_{i=1}^{n-1} (-1)^{\proj[i]{\hat{\bm{y}}}{B}(\left[\bm{z}+\bB_i\right])} \cdot \bm{L}(\bm{z} + \bm{z}_i),\, \forall \bm{z}\in\bE,
\end{equation}
where $\bm{z}_i \in \bB_i$, $\bm{z}_i\neq \bm{0}$ and $\left[\bm{z}+\bB_i\right]$ denotes the coset of $\bE/\bB_i$ containing $\bm{z}$.
Finally, the mean is calculated across all reconstructions and the vector $\bm{L}$ will be updated to $\hat{\bm{L}}$.

These three steps are performed iteratively on each level of recursion until the algorithm converges to a fixed point or a maximum number of $N_{\text{max}}=\lceil m/2\rceil$ iterations has been reached.
The algorithm is said to be converged, if ${|\hat{\bm{L}}(\bm{z}) - \bm{L}(\bm{z})| \leq \theta \cdot |\bm{L}(\bm{z})|}$ for all $\bm{z}\in\bE$, dependent on the threshold $\theta \in\bR_{\geq 0}$.

\subsection{Sparse Recursive Projection-Aggregation Decoding}\label{subsec:srpa}
Sparse RPA (SRPA) \cite{9517887} reduces the complexity of RPA decoding by pruning the projections.
In each recursion, only a random subset ${S=\{s_{1},s_{2},\ldots,s_{p}\} \subseteq \{\bB_1,\bB_2,\ldots,\bB_{n-1}\}}$ of $0<p\leq n-1$ subspaces is used to project the received codeword onto its cosets.
Therefore the aggregation function \eqref{eq:rpa-awgn-prj} must be adapted, leading to \cite{9517887}:
\begin{equation}\label{eq:srpa-awgn-agg}
    \hat{\bm{L}}(\bm{z}) = \frac{1}{p} \sum_{\bB_j\in S} (-1)^{\proj[j]{\hat{\bm{y}}}{B}(\left[\bm{z}+\bB_j\right])} \cdot \bm{L}(\bm{z} + \bm{z}_j),\, \forall \bm{z}\in\bE,
\end{equation}
with $S=\{s_{1},\ldots,s_{p}\}$.
This results in a reduction of projections per recursion by a factor of $(n-1)/p$, reducing $p$ degrades the WER.
If at least 50\% of the projections are used, the loss compared to RPA is negligible.
For more intensive pruning, the WER becomes significantly worse.
To counteract this, sparse multi-decoder RPA ($k$-SRPA) can be used to improve the performance, comparable to a list-decoding approach.
The complexity is then increased again by a factor $k\in\bN$.

\section{Strategic Selection of Subspaces}\label{sec:rpa-extension}

\subsection{Insights to SRPA decoding}\label{subsec:srpa-insights}
As mentioned earlier, the complexity of FHT decoding is $\mathcal{O}(n \log n)$ and thus dominates the complexity of the RPA algorithm.
In the worst case, if all iterations of RPA are carried out (full-RPA), a maximum of
\begin{IEEEeqnarray}{rCl}\label{eq:n-rpa-fhts}
    N_{\text{RPA}} &=& \prod_{i=1}^{r-1} \left\lceil \frac{m-i-1}{2} \right\rceil \cdot \big( 2^{m-i-1} - 1\big)
\end{IEEEeqnarray}
FHT decodings have to be performed.

\begin{definition}[Pruning Factor]
    For the SRPA, we define the reciprocal of the projection reduction factor $(n-1)/p$ as the \textit{pruning factor} $r_p$, where $0 < r_p\leq 1$ and $p$ can be calculated as:
    \begin{equation}\label{eq:pruning-factor}
        p = \big\lceil r_p \cdot (n-1) \big\rceil.
    \end{equation}
\end{definition}
Taking into account the pruning factor $r_p$ in \eqref{eq:pruning-factor} gives:
\begin{equation}\label{eq:n-srpa-fhts}
    N_{\text{SRPA}} = \prod_{i=1}^{r-1} \left\lceil \frac{m-i-1}{2} \right\rceil \cdot \Big\lceil r_p \cdot \big( 2^{m-i-1} - 1\big) \Big\rceil.
\end{equation}
With a pruning factor $r_\text{p}=1/8$ and a third-order RM code, for example, the number of SRPA projections is reduced by up to 93.75\% when compared to full-RPA.

To improve the performance of such low-complexity SRPA decoders, we focus on the choice of the $p$ subspaces used in \eqref{eq:srpa-awgn-agg}.
To find out how the choice of subspaces affects the decoding result, we first take a look at RPA for the binary symmetric channel (BSC).
For an $\RM$ code of length $n$ with the codewords $\bm{c}\in\bF_2^n$, the BSC outputs are $\bm{y}\in\bF_2^n$.
Assuming an SRPA decoder with $|S|=p$ subspaces in each recursion, there exist $\binom{n-1}{p}$ subsets ${ S_{i}\subseteq \{\bB_1,\ldots,\bB_{n-1}\} }$, where $S_{i} \neq S_{j} \, \forall \,i,j\in \{1,\ldots,n-1\}$ and $i\neq j$.
We could now decode the received vectors $\bm{y}$ using all $\binom{n-1}{p}$ SRPA decoders with fixed subsets $S_i$ and see which SRPA, i.\,e., subset of subspaces, performs best on average.
In a sandbox environment, we decoded all possible $\bm{y}\in\bF_2^{16}$ received vectors for an $\rm{4}{2}$ code of length $n=16$ with $\binom{15}{3}=455$ different SRPA decoders using subsets $S_i$ of size $p=3$.
As a result, all decoders lead to the exactly same number of word errors due to the symmetry of the RM code.
Consequently, there is no subset $S_i$ which performs better than any other subset on average.
Albeit, if we are looking at a specific received vector $\bm{y}$, some SRPA decoders decode to the correct $\hat{\bm{c}}= \bm{c}$ while the majority of decoders produce a word error $\hat{\bm{c}}\neq \bm{c}$.
The decoding result therefore depends on the underlying noise pattern.
With the knowledge of the error position, an SRPA decoder with a suitable subset of subspaces could be selected.
Unfortunately we do not have this knowledge for the BSC.

\subsection{Deterministic Subspace Selection Strategy}\label{subsec:sdss-srpa}
Based on the analysis of SRPA decoding given in \ref{subsec:srpa-insights}, we now derive a method to select subspaces for transmission over a general BMS channel.
For an AWGN channel with BPSK modulation (also used in \cite{9023389, 9517887}), the evidence of a received symbol $y\in\bR$ depends on the absolute value of the LLR.
We make use of this fact to select subspaces whose quotient space contains cosets that beneficially combine LLRs indexed by these cosets using \eqref{eq:rpa-awgn-prj}.
We call a subspace $\bB_i$ beneficial if the sum of differences between the absolute LLRs indexed by the vectors $\bm{z}^{(0)}, \bm{z}^{(1)}\in T$ with $T \in \bE/\bB_i$, is small. We denote this distance as $d_{L,\bB_i}$ and define it as:
\begin{IEEEeqnarray}{rCl}\label{eq:llr-distance}
    d_{L,\bB_i} &\coloneqq& \sum_{T \in \bE /\bB_i} d_{\text{LLR}}(\bm{L},T)\nonumber\\
    &=& \sum_{T \in \bE /\bB_i} \Big|\big|\bm{L}(\bm{z}^{(0)})\big|-\big|\bm{L}(\bm{z}^{(1)})\big|\Big|.
\end{IEEEeqnarray}
This distance is based on our investigation of the behavior of SRPA decoders for the $\rm{4}{2}$ code, initialized with all possible subsets of subspaces of size $p=4$, following the approach detailed in Sec.\,\ref{subsec:srpa-insights}.
Without loss of generality, we simulated $10^6$ transmissions of the all-zero codeword over an AWGN channel.
It turned out that decoders with subsets of subspaces that combine LLRs with similar absolute values produce fewer decoding errors.
This behavior can be explained by the fact that the projections in the FHT are correlated with Hadamard sequences.
If two absolute-wise large and therefore reliable LLRs are combined, the corresponding coordinate of the projection has a larger influence on the correlation in the FHT.
The combination of small LLRs, which are potentially more likely to be transmission errors, therefore has less influence on the reconstruction of the projection.
The combination of small and large LLRs potentially leads to a sign change of LLRs that are actually correct and is therefore unfavorable.
Based on \eqref{eq:llr-distance}, we propose a method for deterministic subspace selection (DSS):
\begin{equation}\label{eq:dss}
    \Tilde{S} = \arg \underset{S\in \bS}{\min} \sum_{\bB_i\in S} d_{L,\bB_i},
\end{equation}
where $S=\{s_1,s_2,\ldots ,s_p \}$ is an element of the set $\bS$ containing all $\binom{n-1}{p}$ subsets that are $p$-combinations (without repetition) of subspaces $\bB_1,\bB_2,\ldots,\bB_{n-1}$.
We select the subset $\Tilde{S}$ which has the smallest sum of distances $d_{L,\bB_i}$ over $p$ out of $n-1$ subspaces $\bB_i$.
By calculating the distances $d_{L,\bB_i}$ of all subspaces and choosing the subspaces with the $p$ smallest distances, $\Tilde{S}$ can be determined efficiently.
Our simulations show that this approach works well for second-order RM codes up to length of $n\leq 32$.
For higher-order RM codes and particular for longer RM codes, we present a semi-deterministic subspace selection.

\subsection{Semi-Deterministic Subspace Selection for SRPA Decoding}\label{subsec:sdss}
When we apply \eqref{eq:dss} on longer RM codes with $m=6,7,8,\ldots$, the WER degrades.
In the $N_\text{max}$ iterations of RPA performed on each level of recursion, the subset $\Tilde{S}$ often contains the same subspaces in successive iterations, due to merely small changes in $\hat{\bm{L}}$.
This leads to a saturation effect, where most of the coordinates of $\hat{\bm{L}}$ have the correct sign, but some sign changes in $\hat{\bm{L}}$ cannot be corrected.
This is evident in a low bit error rate (BER) despite a high WER.
To solve this problem, we introduce a \textit{deterministic subspace selection factor} $r_q$, that allows us to vary the grade of determination/randomness in the subspace selection.
\begin{definition}[Deterministic Subspace Selection Factor]
    The deterministic subspace selection factor $r_q \in [0,1]$ is defined as:
    \begin{equation}\label{eq:dss-factor}
        q = \big\lceil (1 - r_q + r_q r_p) \cdot (n-1) \big\rceil,
    \end{equation}
    where $0< p \leq q \leq n-1$.
\end{definition}
Instead of selecting the subset $\Tilde{S}$ containing $p$ subspaces deterministically, we now select $q$ subspaces deterministically according to the same principle, i.\,e., using \eqref{eq:dss} and then choosing $p$ out of $q$ subspaces at random.
If $r_q=0$ the selection is fully random, i.\,e., equivalent to SRPA, and if $r_q=1$ the selection is fully deterministic.
We call the method proposed in this section the \textit{semi-deterministic subspace selection} (SDSS) SRPA.
The SDSS algorithm \texttt{SDSS\_SRPA} is given in Algorithm~\ref{alg:sdss-srpa}.

\begin{algorithm}
\caption{The \texttt{SDSS\_SRPA} Semi-Deterministic Subspace Selection Function for SRPA Decoder} \label{alg:sdss-srpa}
\textbf{Input:} LLR vector $\bm{L}(\bm{z})$; parameter $m$ of the RM code; pruning factor $r_p$; SDSS factor $r_q$

\textbf{Output:} $S^*$ semi-deterministic selected subset of subspaces

\begin{algorithmic}[1]
\State $n \gets 2^{m}$
\State $p \gets \lceil r_p \cdot (n-1) \rceil$ \Comment{number of rand. subspaces}
\State $q \gets \lceil (1 - r_q - r_q r_p) \cdot (n-1) \rceil$ \Comment{number of deterministically selected subspaces}
\State Initialize $\bm{d}_{L,\bB} = (d_{L,\bB_1},d_{L,\bB_2},\ldots,d_{L,\bB_{n-1}}) \in \bR^{n-1}$ as an all-zero vector
\For{$i=1,2,\ldots,n-1$}
    \State $T \gets \bE/\bB_i$ \Comment{$n/2$ unique cosets in each $\bE/\bB_i$}
    \For{$l=1,2,...,n/2$}
        \State $d_{L,\bB_i} \gets d_{L,\bB_i} + \Tilde{d}_\text{LLR}(\bm{L}, T_{l})$ \Comment{LLR differences}
    \EndFor
\EndFor
\State $\Tilde{\bm{d}}_{L,\bB} \gets \texttt{SORT\_ASCENDING}(\bm{d}_{L,\bB})$
\State $\Tilde{S} \gets \arg \big\{ \Tilde{d}_{L,\bB_1},\ldots,\Tilde{d}_{L,\bB_q} \big\}$ \Comment{Select $q$ subspaces} %
\State $S^* \gets \texttt{RAND}(\Tilde{S}, p)$ \Comment{Randomly choose $p$ subspaces}
\State \textbf{return} $S^*$
\end{algorithmic}
\end{algorithm}

\subsection{Optimization of SDSS}\label{subsec:param-optim}
To get the optimal DSS factor $r_q$ in \eqref{eq:dss-factor}, depending on the used pruning factor $r_p$ and the normalized signal-to-noise ratio \SNR, we sweep trough different factors $r_q$ from 0 to 1 and take the one yielding the lowest WER.

Furthermore, we found that a normalization of the distance $d_\text{LLR}(\bm{L}, \bm{z} \in T)$ introduced in \eqref{eq:llr-distance} and a non-linear weighting of the LLRs leads to better results.
An improved distance is defined as:
\begin{equation}
   \Tilde{d}_\text{LLR}(\bm{L},T)  = \bigg|\exp\Big(-\big|\bm{L}(\bm{z}_T^{(0)})\big|\Big) - \exp\Big(-\big|\bm{L}(\bm{z}_T^{(1)})\big|\Big)\bigg|\nonumber
\end{equation}
with $\Tilde{d}_\text{LLR}(\bm{L},T)\in [0,1)$.
Thus, the adapted DSS function is:
\begin{equation}\label{eq:dss-adapted}
    \Tilde{S} = \arg\underset{S\in\bS}{\min}\sum_{i\in S}\sum_{T \in \bE /\bB_i} \Tilde{d}_\text{LLR}(\bm{L},T),
\end{equation}
where $\Tilde{S}=\{s_1,s_2,\ldots ,s_q \}$ is an element of the set $\bS$ containing all $\binom{n-1}{q}$ subsets that are $q$-combinations (without repetition) of subspaces $\bB_1,\bB_2,\ldots,\bB_{n-1}$.
From $|\Tilde{S}|=q$, a subset $S^* \subseteq \Tilde{S}$ with $|S^*|=p$ is chosen at random.
The subspace selection function \eqref{eq:dss-adapted} is likely still not optimal, but optimizing the function directly is combinatorically hard.
For example, there exist $\binom{127}{16}\approx 10^{19}$ different subsets $S$ for an \rm{7}{2} code of length $n=128$ and a pruning factor of $r_p=1/8$.
This justifies the independent selection of individual subspaces $\bB_i$ based on our proposed methods, instead of looking at the whole set of subsets $S$.

In addition to SRPA decoding, we perform Reed's decoding on the reconstruction at the end to ensure that the decoding result is a valid codeword.
Since Reed's decoding is a low-complexity decoder of complexity $\mathcal{O}(n \log^r n)$, the overhead is negligible compared to the complexity of SRPA.
This is also done in \cite{9023389,9517887}.

\begin{figure*}[ht]
    \centering
    \input{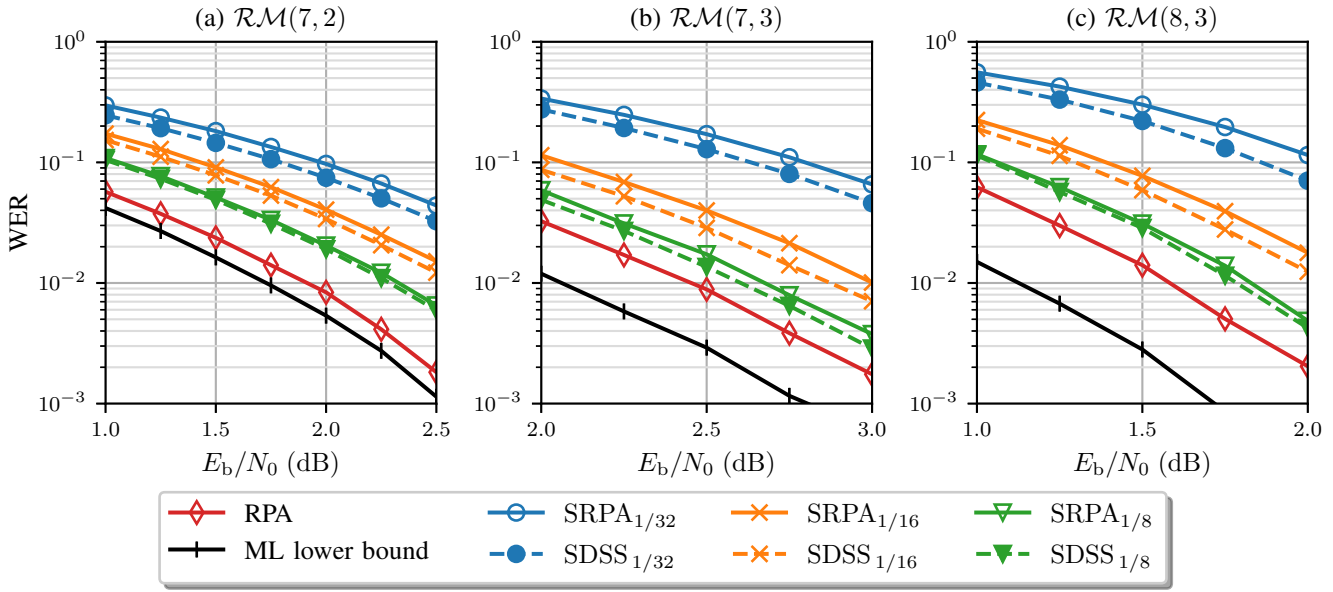}
    \caption{Simulated transmissions over an AWGN channel for second- and third-order RM codes decoded with SRPA and SDSS-SRPA decoders and different pruning factors $r_p$, indicated as $\text{SRPA}_{r_p}$ and $\text{SDSS}_{r_p}$.}
    \label{fig:rm-codes-results}
\end{figure*}

\subsection{Complexity Reduction of SDSS-SRPA}\label{subsec:complexity-reduction}
Additionally to the introduced subspace selection method, which improves the performance of SRPA decoding, the complexity can be further reduced.
Actually, it is not necessary to perform $N_\text{max}$ iterations at each level of the $r-1$ recursions.
The iterations on lower levels up to the $\rm{m-r+1}{1}$ code have little impact on the decoding result, but on the other hand increase the decoding complexity significantly.
Hence, we only perform the first recursion iterations and therefore reduce the maximum number of FHTs given in \eqref{eq:n-srpa-fhts} for the SRPA to:
\begin{equation}\label{eq:n-sdss-fhts}
    N_{\text{SDSS}} = \left\lceil \frac{m}{2} \right\rceil \prod_{i=1}^{r-1} \Big\lceil r_q \cdot \big( 2^{m-i-1} - 1\big) \Big\rceil.
\end{equation}
The most important iterations are those at the top level before the first projection, i.e., for the \rm{m}{r} code.
This observation has very recently also be observed in \cite{9414655, 2210.11069}, but without the gain in the decoding performance we achieve from SDSS.
At the top level, two iterations are usually sufficient for the algorithm to converge.
Therefore we use the early stopping technique introduced for RPA with the threshold $\theta$ for our novel SDSS-SRPA decoding method.
Consequently, the average number of FHTs performed can be even lower than upper bound \eqref{eq:n-sdss-fhts}.
To measure the complexity reduction, we define the complexity reduction gain as follows:
\begin{equation}\label{eq:c-reduction-gain}
    G_\text{C} \coloneqq 1 - \overline{N}_\text{SDSS} / N_\text{SRPA},
\end{equation}
where $\overline{N}_\text{SDSS}$ denotes the average number of first-order RM code FHT decodings over $N \in \bN$ decoded received codewords including the early stopping criterion.

\section{Results}\label{sec:results}
The SDSS extension is tested on second- and third-order RM codes of length $n\leq 256$ with single-decoder SRPA and pruning factors $r_p\leq 1/2$.
Unlike \cite{9517887}, no CRC polynomial is used, since it reduces the effective code rate.
The SDSS method with the additional early stopping criterion and omitted inner iterations is compared to SRPA decoding with the same pruning factor. %
Additionally, these results are compared to RPA decoding and the lower bound on the performance of ML decoding described in \cite{1603792, 1291729}.
The simulation results are shown in Fig.\,\ref{fig:rm-codes-results}, where at least $10^5$ codewords are simulated for each $\SNR$.

The performance gain of the SDSS method compared to SRPA varies between $0.01\,\text{dB}$ to $0.2\,\text{dB}$.
The largest gains are achieved for small pruning factors where only few projections are used.
For a pruning factor of $r_p=1/32$, a gain by up to $0.2\,\text{dB}$ can be observed.
For larger pruning factors, the $\SNR$-gain becomes smaller.
This is due to the random choice of subspaces by the SRPA.
The more subspaces are used for the projection, the more likely it is that a random subset of subspaces also contains those subspaces that are selected by the SDSS method.
Although a gain can be obtained through SDSS, the performance of SRPA for small pruning factors $r_p$ is no longer as close to ML decoding as RPA is.
However, if a low complexity decoder is required, e.\,g., for hardware implementation, the SDSS method significantly improves the performance of the SRPA decoder with small pruning factors while at the same time reducing the computational burden.

In addition to the $\SNR$ gain of the SDSS method mentioned before, a complexity reduction gain $G_\text{C}$, defined in \eqref{eq:c-reduction-gain}, can by achieved by the methods introduced in Sec.\,\ref{subsec:complexity-reduction}.
Nevertheless, the deterministic subspace selection requires some additional computations when compared to the random selection in SRPA.
In the implementation, the SDSS is carried out efficiently during the projection step of the RPA algorithm.
Overall, for second-order RM codes, the runtime of both algorithms is almost identical in our simulations or even slightly lower for the SDSS method, due to the early stopping criterion.
In most cases, two iterations are sufficient for the algorithm to converge.
Thus, the complexity reduction gain $G_\text{C}$ caused by the early stopping criterion, becomes larger if long RM codes are used, since the maximum number of iterations $N_\text{max}$ depends on the parameter $m$.
Furthermore, a higher $\SNR$ results in a faster convergence and thus in a complexity reduction.

However, the largest complexity gains $G_\text{C}$ are achieved if RM codes of order $r\geq 3$ are used and the lower recursion level iterations are omitted.
For third-order RM codes, this leads to a complexity gain $G_\text{C}$ by up to 81\%, as it is shown in Tab.\,\ref{table:num-fhts}.
This results in a nearly five-times faster decoding compared to SRPA.
For higher-order RM codes, $G_\text{C}$ would be even larger.
\begin{table}[t!]
\caption{Number of FHT Decodings -- SRPA vs. SDSS}
\begin{center}
\begin{tabular}{cccccccc}
\toprule
\multirow{2}{*}{$r_p$}& \multicolumn{3}{c}{$\rm{7}{3}$} &  & \multicolumn{3}{c}{$\rm{8}{3}$}\Tstrut\Bstrut\\
\cline{2-4} \cline{6-8}
             & $N_\text{SRPA}$ & $N_\text{SDSS}$ & $G_\text{C}$   &  & $N_\text{SRPA}$ & $N_\text{SDSS}$ & $G_\text{C}$ \Tstrut\Bstrut\\ \midrule
1/16       & 384     & 128   & 67\%    &  & 2048      & 511     & 75\%\Tstrut\\
1/8        & 1536    & 510     & 67\%    &  & 8192      & 1981    & 76\%\\
1/4        & 6144    & 1805    & 71\%    &  & 32768     & 7126    & 78\%\\
1/2        & 24576   & 6267    & 74\%    &  & 131072    & 25046   & 81\%\Bstrut\\
\bottomrule
\end{tabular}
\label{table:num-fhts}
\end{center}
\end{table}

\begin{figure}[hb!]
    \centering
    \resizebox{1.0\linewidth}{!}{\input{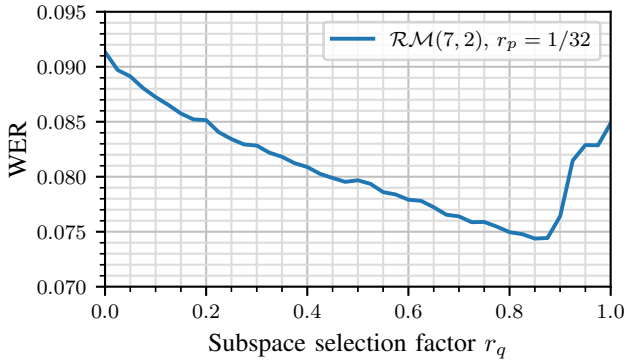}}
    \caption{Optimization of the DSS factor $r_q$ for the \rm{7}{2} code with pruning factor $r_p=1/32$ and $\SNR = 2\,\text{dB}$. The optimum DSS factor equals $r_q=0.85$ at $\text{WER}\approx7.45\cdot 10^{-2}$.}
    \label{fig:optim}
\end{figure}

The DSS factor $r_q$ of the SDSS method is optimized for all simulated RM codes and each pruning factor.
The optimal $r_q$ depends on the $\SNR$, but is almost constant for the $\SNR$-ranges shown in Fig.\,\ref{fig:optim}.
For this reason, $r_q$ is optimized for $\SNR = 2\, \text{dB}$ and henceforth assumed to be independent of \SNR.
Figure~\ref{fig:optim} shows the optimization result for the \rm{7}{2} code, $r_p=1/32$ and $\SNR = 2\,\text{dB}$ as an example.
In this case, the optimal subspace selection factor yielding the lowest WER of $7.45\cdot 10^{-2}$ is $r_q=0.85$.

\section{Conclusion}\label{sec:conclusion}
The performance of SRPA decoding of RM codes can be improved using the methods proposed in this paper to select the subspaces used for the projections in a semi-deterministical matter, depending on the SRPA pruning factor.
This subspace selection method is based on a figure of merit that evaluates differences between the LLR values indexed by all cosets.
In addition, the computational burden is significantly reduced while obtaining a performance gain in the order of $0.01\,\text{--}\,0.2\,\text{dB}$.
This complexity reduction is achieved by performing fewer iterations, and thus, reducing the number of ML decodings of first-order RM codes.
Optimizing the subspace selection function is subject to future work, potentially further improving the performance of low complexity SRPA decoding of RM codes.

\bibliographystyle{IEEEtran}
\bibliography{bibliography-short}

\end{document}